\documentclass[11pt]{article}
\usepackage{cite}
\usepackage{tikz}
\usetikzlibrary{arrows,shapes}
\usetikzlibrary{trees}
\usetikzlibrary{matrix,arrows} 				
\usetikzlibrary{positioning}				
\usetikzlibrary{calc,through}				
\usetikzlibrary{decorations.pathreplacing}  
\usepackage{pgffor}							

\usetikzlibrary{decorations.pathmorphing}	
\usetikzlibrary{decorations.markings}
\tikzset{
	vector/.style={decorate, decoration={snake}, draw},
	provector/.style={decorate, decoration={snake,amplitude=2.5pt}, draw},
	antivector/.style={decorate, decoration={snake,amplitude=-2.5pt}, draw},
	fermion/.style={draw=black, postaction={decorate},
		decoration={markings,mark=at position .55 with {\arrow[draw=black]{>}}}},
	fermionbar/.style={draw=black, postaction={decorate},
		decoration={markings,mark=at position .55 with {\arrow[draw=black]{<}}}},
	fermionnoarrow/.style={draw=black},
	gluon/.style={decorate, draw=black,
		decoration={coil,amplitude=4pt, segment length=5pt}},
	scalar/.style={dashed,draw=black, postaction={decorate},
		decoration={markings,mark=at position .55 with {\arrow[draw=black]{>}}}},
	scalarbar/.style={dashed,draw=black, postaction={decorate},
		decoration={markings,mark=at position .55 with {\arrow[draw=black]{<}}}},
	scalarnoarrow/.style={dashed,draw=black},
	electron/.style={draw=black, postaction={decorate},
		decoration={markings,mark=at position .55 with {\arrow[draw=black]{>}}}},
	bigvector/.style={decorate, decoration={snake,amplitude=4pt}, draw},
}

\tikzstyle{block} = [draw, rectangle, 
minimum height=3em, minimum width=6em]

\usepackage{amsmath,amsfonts,amssymb}
\usepackage{graphicx,rotate,multicol,wrapfig}
\usepackage{color}
\usepackage{epsfig,epsf}
\usepackage{bm}
\usepackage{dcolumn,subfig}
\usepackage{array}
\newcolumntype{C}[1]{>{\centering\let\newline\\\arraybackslash\hspace{0pt}}m{#1}}

\usepackage{relsize}
\def\Babar{{\mbox{\slshape B\kern-0.1em{\smaller A}\kern-0.1em B\kern-0.1em{\smaller A\kern-0.2em R}}}}

\usepackage[bookmarks, breaklinks, colorlinks,urlcolor=blue, citecolor=red, 
linkcolor=blue]{hyperref}

 \usepackage{color}

 \usepackage[normalem]{ulem}

 \definecolor{darkgreen}{cmyk}{1,0,1,0.4}
 
 \definecolor{pink}{cmyk}{0.4,1,0.3,0}

\def\com2#1{\textcolor{red}{\it{#1}}}

\def\a {\alpha}

\def\g {\gamma}
\def\d {\delta}

\def\bar {\overline}

\def\lamsb {V_{ts}^\ast V_{tb}}
\def\lamsbtp {V_{t's}^\ast V_{t'b}}
\def\lsbtp {\lambda_{sb}^{t'}}

\def\r {\rightarrow}

\def\n {\nonumber}

\mathchardef\hyp="2D

\def\bbbar{B^0\, \hyp\, \overline{B}^0}

\def\bsbsbar {B_s\, \hyp\, \overline{B_s}}
\def\bqbqbar {B_q\, \hyp\, \overline{B_q}}
\def\zbb {Z\to b \bar{b}}
\def\bsmumu {B_s \to \mu^+\mu^-}
\def\bsg {b \to s\gamma}

\def\mmu {m_\mu}
\def\mw {M_W}
\def\mz {M_Z}

\def\mbs {M_{B_s}}
\def\mt {m_t}

\def\sw {s_W}

\def\g{\gamma}

\def\beq{\begin{equation}}
\def\eeq{\end{equation}}
\def\bea{\begin{eqnarray}}
\def\eea{\end{eqnarray}}
\def\barr{\begin{array}}
\def\earr{\end{array}}

\def\abs#1{\left| #1 \right|}

\def\gev{\ensuremath{\mathrm{Ge\kern -0.1em V}}}

\begin{document}

\renewcommand*{\thefootnote}{\fnsymbol{footnote}}

\begin{center}
{\Large\textbf{Constraints on the quark mixing matrix with vector-like quarks}}\\[3mm]
{\large Drona Vatsyayan}\footnote{E-mail: dronavatsyayan@gmail.com} 
and {\large Anirban Kundu}\footnote{E-mail: anirban.kundu.cu@gmail.com}\\[2mm]
{\em\small Department of Physics, University of Calcutta,\\
92 Acharya Prafulla Chandra Road, Kolkata 700009, India}
\end{center}

\begin{abstract}

While a chiral fourth generation of quarks is almost ruled out from the data on Higgs boson production and 
decay at the Large Hadron Collider, vector-like quarks are still a feasible option to extend the fermionic 
sector of the Standard Model. Such an extension does not suffer from any anomalies and easily passes 
the constraints coming from oblique electroweak parameters. We consider such minimal extensions with $SU(2)$ 
singlet and doublet vector-like quarks that may mix with one, or at the most two, of the Standard Model quarks.
Constraints on the new mixing angles and phases are obtained from several $\Delta B = 1$ and 
$\Delta B = 2$ processes. 

\end{abstract}

\date{\today}

\setcounter{footnote}{0}
\renewcommand*{\thefootnote}{\arabic{footnote}}

\section{Introduction}

While all flavour observables, both CP-conserving and CP-violating, are more or less well explained by the 
$3\times 3$ Cabibbo-Kobayashi-Maskawa (CKM) matrix, it is pertinent to investigate whether one can still 
accommodate one or more extra quarks. The new quarks can be chiral, like their Standard Model (SM) 
counterparts, or vector-like, where both left- and right-chiral components transform identically under the 
$SU(2)$ of weak interaction. A chiral fourth generation \cite{He:2001tp,Kribs:2007nz}
that gets its mass through the same Higgs mechanism as the other three generations do, is more or less 
ruled out from the induced oblique corrections \cite{Erler:2004nh} and the
data on Higgs boson production and decay at the Large Hadron Collider (LHC) \cite{Chen:2012wz,Eberhardt:2012gv,Djouadi:2012ae,Kuflik:2012ai}, 
unless the scalar sector of the Standard Model (SM) is extended \cite{Das:2017mnu}. However, vector-like quarks 
(VQ) easily evade such a constraint \cite{Lavoura:1992np}. Moreover, by construction, inclusion of VQs still keeps the SM free
from any gauge or mixed anomalies. 

The VQs appear naturally in many extensions of the SM, and their effects and signatures have been 
thoroughly investigated. An overview of the phenomenology and searches for heavy VQs has been provided in Ref.\ \cite{Okada:2012gy}. For example, the $SU(2)$ doublet VQs in the topflavour seesaw models, and their 
signatures at the LHC, have been discussed in Refs.\ \cite{He:1999vp,Wang:2013jwa}, or the role of the VQs in Higgs inflationary scenarios in Ref.\ \cite{He:2014ora}. For the study of nonunitarity of the CKM matrix, or the implications 
of a singlet down-type VQ, and its implications at the LHC, we refer the reader to Refs.\ 
\cite{Liao:2000re,Lavoura:1992qd,Higuchi:2009dp,Karabacak:2014nca}. The radiative decays of 
the $b$-hadrons were discussed in Refs.\ \cite{Chang:2000gz,Aoki:2000ze,Akeroyd:2001gf}, while the
role of VQs in $\bsbsbar$ mixing was investigated in Refs.\ \cite{Barenboim:1997pf,AguilarSaavedra:2002kr,Aoki:2001xr,Barenboim:2001fd}, and that in semileptonic and hadronic $B$-decays in Ref.\ \cite{Morozumi:2004ea}.

In this work, we would be interested to find out how much mixing any new VQ may have with the SM 
quarks. For simplicity, we consider only one charge $+\frac23$ VQ, called $t'$, 
and/or only one charge $-\frac13$ VQ, called $b'$. (Thus, SM gauge singlet VQs, a possible fermionic 
dark matter candidate, are outside our purview.) 
In other words, we would like to extend the CKM matrix by a row, or a column, or both, and try to find 
the bound on the new elements. Such studies have been performed 
earlier \cite{Eberhardt:2010bm,AguilarSaavedra:2002kr,Botella:2012ju,Alok:2014yua,Alok:2015iha}, 
and we will build on those studies, armed with more precise experimental data. Our main tool will be the loop-mediated $\Delta B=1$ ($\bsg$, $\bsmumu$) or $\Delta B = 2$ ($\bsbsbar$ mixing) 
processes, and occasionally, the decay $\zbb$. This is one of the reasons 
why we will not discuss the chiral fourth generation with more scalar fields, as those
scalars affect these observables in a nontrivial way. A good example is the radiative $b$ decay, $\bsg$, 
which is affected by the charged scalar loop. Thus, the bounds on the new CKM elements are meaningful 
only if we have a complete control over the parameters of the scalar potential. As a spin-off, we will also
show how anomalous top decays like $t\to c\gamma$ and $t\r cZ$ are affected. 

If a chiral quark is extremely massive, say of the order of a TeV, the Yukawa coupling will be badly 
nonperturbative, and no loop calculations are trustworthy \cite{Denner:2011vt}. For VQs, one does not face such 
a problem, as they can have a mass term without symmetry breaking. No such heavy quarks have been 
observed, so for our analysis, we will keep their masses fixed at 2 TeV. This is just outside the LHC reach; 
however, in future, only the luminosity of LHC, and not the energy reach, will increase, so anything 
beyond 2 TeV should remain undiscovered. This increases the importance of such indirect detection 
studies.  


The paper is organised as follows. In Section \ref{sec:sm}, we provide a short introduction to the VQ models 
as well as a brief summary of the SM expressions for various loop-mediated observables that are relevant for our 
study. We also enlist the data that will be used in our analysis. 
In Section \ref{sec:expvlq}, we recast these expressions in the framework of the extended SM with 
vector-like singlet or doublet quarks. In Section \ref{sec:results}, we display our results for each of these 
models, followed by the summary and conclusion in Section \ref{sec:summary}.

\section{The toolbox}  \label{sec:sm}

The charged current Lagrangian in the SM involving the stationary quark fields looks like
\beq
{\cal L}_{cc} = -\frac{g}{\sqrt{2}} \, \bar{u}_i V_{ij} \gamma^\mu P_L d_j W^+_\mu  + {\rm h.c.}
\eeq
where $i,j=1,2,3$ are the generation indices, $P_L = \frac12(1-\gamma_5)$,  and $V$ is the CKM matrix.
By construction, $V$ is unitary, and the magnitudes of seven of the elements, as determined from tree-level 
processes, are as follows:  
\begin{table}[!htb]
    \centering
    \begin{tabular}{||c|c||c|c||c|c||}
    \hline
       Element & Value &  Element & Value & Element & Value \\[0.5ex]
       \hline \hline
       $V_{ud}$ & $0.97370 \pm 0.00014$ & 
       $V_{us}$ & $0.2245 \pm 0.0008$ & 
       $V_{ub}$ & $(3.82 \pm 0.24) \times 10^{-3}$\\
       $V_{cd}$ & $0.221 \pm 0.004$ & 
       $V_{cs}$ & $0.987 \pm 0.011$ & 
       $V_{cb}$ & $(41.0 \pm 1.4) \times 10^{-3}$\\
       & & & & $V_{tb}$ & $1.013 \pm 0.030$ \\
       \hline
    \end{tabular}
    \caption{Magnitudes of the directly measured CKM matrix elements, taken from Ref.\ \cite{pdg}.}
    \label{tab:a}
\end{table}

The elements $V_{td}$ and $V_{ts}$ can only be determined from $\bbbar$ or $\bsbsbar$ mass difference,
or indirectly from the unitarity of the CKM matrix. In our analysis, we will implicitly assume that the seven 
CKM elements shown in Table \ref{tab:a} are not affected in any significant way by the VQs. In particular, 
we will strictly assume that the mixing of the VQs with the first generation vanishes, so that processes like
$K$-decays remain unaffected. 

\subsection{Introducing the vector-like quarks}   \label{sec:vq}

Let us first introduce a charge $-\frac13$ VQ, $b'$, which is an $SU(2)$ singlet. 
The charged current Lagrangian now reads
\beq
{\cal L}_{cc} = -\frac{g}{\sqrt{2}} \, \bar{u}_i\, {\cal U}^\dag_{ij} {\cal D}_{jk}\, \gamma^\mu P_L\, 
d_k W^+_\mu  + {\rm h.c.}
\label{eq:lcc}
\eeq
where ${\cal U}$ and ${\cal D}$ are the $3\times 3$ and $4\times 4$ basis transformation matrices. 
The indices can have the following values: $i,j = 1,2,3$, and $k = 1,2,3,4$.  
Hence, the CKM matrix $V \equiv {\cal U}^\dag {\cal D}$ is a $3\times 4$ matrix. Obviously, the SM part 
of $V$ is no longer unitary, and we cannot use the unitarity condition to constrain its elements. The 
following points immediately become evident:
\begin{itemize}
\item  
Such a $3\times 4$ CKM matrix can be parametrized by 9 independent parameters, including 6 real
angles and 3 phases. To see this, let us take ${\cal U}={\bf 1}$, the identity matrix, so that $V$ is just the
top $3\times 4$ block of ${\cal D}$. The row unitarity conditions still hold, which gives 9 constraints, 
and one can remove 6 irrelevant phases through quark phase redefinition. Thus, one is left with 
$2\times 12 - 9 - 6 = 9$ independent parameters. 

\item The tree-level $Z$-mediated flavour-changing neutral current (FCNC) processes will be present 
in the down sector, but not in the up sector. The vertex factor for $\bar{d}_{iL} d_{jL} Z$ tree-level  interaction is
\beq
g^Z_{ijL} = \frac{g}{c_W}\, \left( -\frac12 + \frac13 s_W^2\right) \delta_{ij} + \frac{g}{2c_W} {\cal D}^*_{4i}
{\cal D}_{4j}\,,
\label{eq:fcnc1}
\eeq
while there is no such FCNC for the right-chiral quarks (the situation is exactly the opposite for vector-like 
doublet quarks). We use the standard abbreviation of $s_W \equiv \sin\theta_W$, $c_W\equiv\cos\theta_W$. Such FCNC vertices can have important implications in flavour physics, and we may refer the reader to,
{\em e.g.}, Refs.\ \cite{Barenboim:1997qx,Branco:1992wr}, for its effects to the CP
asymmetries in the $\bbbar$ system.

\item
Only if we consider the texture where ${\cal U} = {\bf 1}$ and hence ${\cal D}_{3\times 4} = V$, 
the FCNC vertex factor in Eq.\ (\ref{eq:fcnc1}) can be written as
\beq
g^Z_{ijL}~(i\not= j) = -\frac{g}{2c_W}\, \left[V_{ui}^* V_{uj} + V_{ci}^* V_{cj} + V_{ti}^* V_{tj}\right]\,,
\label{eq:fcnc2}
\eeq
which can potentially be most significant for the $bsZ$ vertex. 

\item
Again, under such assumptions like ${\cal U}={\bf 1}$, the $3\times 4$ CKM matrix is a part of the full 
$4\times 4$ ${\cal D}$ matrix, which is unitary, and we can use the row unitarity conditions as constraints,
{\em e.g.},
\beq
\sum_{i=d,s,b,b'} \left\vert V_{ti}\right\vert^2 = 1\,.
\label{eq:rowunit}
\eeq
\end{itemize}

Similarly, one can introduce a charge $+\frac23$ singlet VQ, $t'$, for which the CKM matrix $V$ is $4\times 3$. 
The FCNC will be present only for left-chiral up-type quarks, parametrized by a vertex factor analogous to
that in Eq.\ (\ref{eq:fcnc1}),
\beq
g^Z_{ijL} = \frac{g}{c_W}\, \left( \frac12 - \frac23 s_W^2\right) \delta_{ij} - \frac{g}{2c_W} {\cal U}^*_{4i}\,
{\cal U}_{4j}\,,
\label{eq:fcnc3}
\eeq
which leads to the tree-level top FCNC decays like $t\to c Z$. 

We will consider the following simplified models for singlet VQ (not all of them can be constrained from 
flavour observables):
\begin{enumerate}
\item {\bf Model VQ-S-D1}: A charge $-\frac13$ VQ $b'$ with $V_{ub'}=V_{cb'} =0$, $V_{tb'}\not= 0$. The structure
forbids any tree-level FCNC. However, there will be new one-loop contributions to $ZWW$ and $\gamma WW$ 
vertices, as well as to $t\bar{t}$ production\footnote{The letters S and D in the model denote that the VQ is 
singlet and down-type.}. As we will show, this model remains essentially unconstrained, except the 
bound on $|V_{tb'}|$ coming from unitarity, see Eq.\ (\ref{eq:rowunit}).

\item {\bf Model VQ-S-D2}: Similar to VQ-S-D1, except that both $V_{tb'}$ and $V_{cb'}$ are taken to be nonzero, 
while $V_{ub'}$ is kept fixed at zero. One of these two new elements can contain a nontrivial phase. 
The tree-level $bsZ$ interaction contributes to $\bsbsbar$ mixing, $\bsg$, and $B_s \to \mu^+\mu^-$. 
There will also be loop-mediated contributions to $t\to c\gamma$ \cite{AguilarSaavedra:2002ns}. To maximise the 
effects of the VQs, we will take ${\cal U}={\bf 1}$, so that the FCNC coupling is given by Eq.\ 
(\ref{eq:fcnc2}). 

\item {\bf Model VQ-S-U1}: A mirror image of VQ-S-D1, with one singlet charge $+\frac23$ quark $t'$, 
and $V_{t'd}=V_{t's}=0$ while $V_{t'b}\not = 0$. The CKM matrix is $4\times 3$. 
The texture affects the triple gauge vertices at one-loop, as well as 
the decay $Z\to b\bar{b}$. Again, the most significant constraint on $V_{t'b}$ comes from the column 
unitarity, analogous to Eq.\ (\ref{eq:rowunit}), under possible assumptions about ${\cal D}$.  

\item {\bf Model VQ-S-U2}: Again, similar to VQ-S-U1, but keeping two nonzero mixing elements $V_{t'b}$ and
$V_{t's}$. The same observables as for VQ-S-D2 are affected, but there is a major difference. For VQ-S-D2, 
processes like $\bsbsbar$ mixing were affected by the tree-level $bsZ$ vertex, while $t\to c\gamma$ was 
affected by one-loop processes (with the $b'$ quark running in the loop). For VQ-S-U2, it is just the opposite:
$t\to c\gamma$ is affected by the tree-level $tcZ$ interaction, while $B$-physics observables are affected 
by one-loop processes.  For the VQ-S-U type models, we will consider ${\cal U}_{3\times 4}= V^\dag$ and 
${\cal D} = {\bf 1}$; this maximises the FCNC effects. 

\end{enumerate}

Next, let us consider a degenerate (so that constraints from oblique parameters \cite{Lavoura:1992np} are
under control) vector-like 
doublet $Q = (t'\quad b')^T$, for which both left- and right-chiral components are in the fundamental representation 
of weak $SU(2)$. Eq.\ (\ref{eq:lcc}) now extends to all the left-chiral quarks, {\em i.e.}, all the indices $i,j,k$ run from 
1 to 4. Obviously, there is no tree-level FCNC in the left-chiral sector. In the right-chiral sector, the FCNC
vertices, for the up- and down-quark sectors respectively, are
\begin{eqnarray}
g^{uZ}_{ijR}~(i\not= j) &=& \frac{g}{2c_W} \left({\cal U}_R\right)^*_{4i}\, \left({\cal U}_R\right)_{4j}\,,\nonumber\\
g^{dZ}_{ijR}~(i\not= j) &=& - \frac{g}{2c_W} \left({\cal D}_R\right)^*_{4i}\, \left({\cal D}_R\right)_{4j}\,,
\label{eq:fcnc4}
\end{eqnarray} 
where ${\cal U}_R$ and ${\cal D}_R$ are the basis transformation matrices for the right-chiral up- and down-type
quarks respectively. The flavour-conserving vertices are exactly like the SM. As the CKM matrix does not contain
any information about ${\cal U}_R$ or ${\cal D}_R$ matrices, one may either treat them as free parameters, 
or just set all the off-diagonal terms to zero ({\em i.e.}, ${\cal U}_R = {\cal D}_R = {\bf 1}$).

One may note that the $4\times 4$ CKM matrix is unitary, so it may be parametrized by 6 real 
angles and 3 complex phases. We do not need to know the individual ${\cal D}_L$ or ${\cal U}_L$ matrices, and 
the analysis is similar to that for a sequential fourth generation \cite{Eberhardt:2010bm}. The unitarity constraints 
can now be invoked, {\em e.g.},
\beq \label{eq:unitarity}
\left\vert V_{t'b}\right\vert^2 = 1 - \sum_{i=u,c,t} \left\vert V_{ib}\right\vert^2\,.
\eeq

We will take ${\cal U}_R = {\cal D}_R = {\bf 1}$, so that for doublet VQ, we only need to consider the loop-driven 
processes. Two models will be discussed, namely:
\begin{enumerate}
\item {\bf Model VQ-D-U2}: Apart from $V_{t'b'}$, only two other elements involving a VQ are taken to be 
nonzero: $V_{t's}$ and $V_{t'b}$. The $b'$ quark decays through a real (if kinematically allowed) or a virtual
$t'$. This texture affects processes like $\bsbsbar$ mixing, $\bsg$, $\bsmumu$, and 
$\zbb$. Another variant is to take $V_{t'd}\not=0$ and $V_{t's}=0$; all the above processes are, in that case, 
replaced by their $s\to d$ analogues. 

\item {\bf Model VQ-D-D2}: The nonzero elements are taken to be $V_{cb'}$ and $V_{tb'}$, apart from 
$V_{t'b'}$. Such a choice affects top pair production, and decays like $t\to c\gamma$. 
\end{enumerate}

One might be wondering about the constraints coming from the oblique parameters. We refer the reader 
to the comprehensive analysis done in Ref.\ \cite{Lavoura:1992np}. This contains the most general case of 
an arbitrary number of singlet and doublet VQs, as well as the specific cases of a down-type or up-type singlet VQ. 
It turns out that for small mixing with chiral quarks (at the level that we are considering), all the models fall
safely within the allowed regions of the oblique parameters.

Being loop-driven in the SM, the amplitudes for all the processes that we consider can be written in a 
generic way:
\beq
{\cal M} = \sum_i A_i {\cal V}_i \, F^{IL}_i (m)\,,
\eeq
where $i$ runs over all the individual amplitudes, including higher order QCD and electroweak corrections,
${\cal V}_i$ is some combination of the CKM 
elements, and $F^{IL}_i(m)$ is the relevant Inami-Lim function, which depends on the quark (mostly top) 
and the gauge boson masses. All the other constants are clubbed in $A_i$.
For our analysis, we will take only the leading order effects coming from the VQs and try to find out
the allowed region of the CKM elements vis-\`a-vis the VQ couplings.

\subsection{\texorpdfstring{$\Delta B = 2$ processes: $\bqbqbar$ mixing}{Delta B=2 processes}} \label{sec:db2}

The mass difference $\Delta M_q$ ($q=d,s$) for the $\bqbqbar$ system is given by
\beq \label{eq:delmq}
\Delta M_q = \frac{G_F^2}{6\pi^2} \, \abs{V_{tq} V_{tb}^\ast}^2 \, \eta_B M_{B_q} (B_{B_q}F_{B_q}^2)\,  
M_W^2 S_0(x_t) \,,
\eeq
where $x_t = m_t^2/M_W^2$, and the Inami-Lim function is 
\beq
S_0(x_t) = \frac{4x_t - 11x_t^2 + x_t^3}{4(1-x_t)^2}-\frac{3 x_t^2\, \log{x_t}}{2(1-x_t)^3}\,,
\eeq
where we neglect the charm- and up-quark contributions. We will focus only on the $q=s$ case. 
All other relevant quantities are shown in Table \ref{tab:sminput}.  

\begin{table}[!htb]
\centering
 \begin{tabular}{||c|c||c|c||} 
 \hline
 Parameter & Value & Parameter & Value \\ [0.5ex] 
 \hline\hline
 $G_F$ & $1.1663787\times 10^{-5}~{\rm GeV}^{-2}$ & $\eta_B$ & $0.55 \pm 0.02$ \\ 
 \hline
 $M_W$ & $80.379 \pm 0.024~{\rm GeV}$ & $m_t$ & $172.4 \pm 1.4~{\rm GeV}$ \\
 \hline
 $M_{B_s}$ & $5366.88 \pm 0.28~{\rm MeV}$ & $\Delta M_s$ & $(1.1683 \pm 0.0026) \times 10^{-8}~{\rm MeV}$ \\ 
 \hline
 $\sqrt{B_{B_s}} F_{B_s}$ & $274 \pm 16~{\rm MeV}$ &  &  \\ 
 %
 %
 %
 \hline
\end{tabular}
\caption{Values of the parameters required for determination of $|V_{ts}|$. The error bars are at 
$2\sigma$  \cite{pdg}.} 
\label{tab:sminput}
\end{table}

This leads to
\bea
 x_t &=& 4.60 \pm 0.04\,,\ \ S_0 (x_t)=2.52 \pm 0.21\,, \nonumber \\
  \Delta M_s &=& (8.3 \pm 0.9) \times 10^{-6}~{\rm MeV} \times |V_{ts}|^2\,
\eea
The phase coming from the mixing in the $\bsbsbar$ system is given by
\beq
M_{12}^s = \frac12\, \Delta M_s \, \exp(-2i\beta_s)\,,
\eeq
where $M_{12}^s$ is the absorptive part of the $(12)$-element of the Hamiltonian matrix, and the 
phase comes from $V_{ts}^2$, whose value is found to be \cite{pdg}
\beq
\beta_s = (1.1\pm 1.6)\times 10^{-2}\,.
\eeq

\subsection{\texorpdfstring{$\Delta B = 1$ processes}{Delta B=1 processes}} \label{sec:db1}

\subsubsection{\texorpdfstring{$\bsg$}{b->s gamma}}

The radiative decay $\bsg$ has a branching ratio of \cite{Misiak:2017woa}
\beq
{\rm Br}(\bsg) = (3.32\pm 0.15)\times 10^{-4}\,,\ \ \ E_\gamma > 1.6~{\rm GeV} 
\eeq
with a corresponding SM prediction of \cite{Misiak:2015xwa,Czakon:2015exa} \footnote{
The SM uncertainties may have been underestimated by about a factor of two, which provides 
more room for VQs and/or any other new physics \cite{Bernlochner:2020jlt}.}
\beq
{\rm Br}(\bsg)_{\rm SM} = (3.36\pm 0.23)\times 10^{-4}\,,\ \ \ E_\gamma > 1.6~{\rm GeV}\,.
\eeq
To reduce the theoretical uncertainties, one generally uses the ratio of the decay rates:
\beq
     R \equiv \frac{\Gamma(B \to X_s \gamma)}{\Gamma(B \to X_c e \bar{\nu}_e)}\simeq 
     \frac{\Gamma(\bsg)}{\Gamma(b \to c e \bar{\nu}_e)}\,.
\eeq
At leading logarithmic order, the ratio can be written as
\beq\label{eq:rbsg}
    R=\frac{|V_{ts}^{*}V_{tb}|^2}{|V_{cb}|^2}\, \frac{6\alpha}{\pi f(z)}\, 
    \left\vert C_{7\gamma}^{(0)\text{eff}}(\mu)\right\vert^2 \,, 
\eeq 
where 
\begin{align}
    f(z)=1-8z^2-8z^6-z^8-24z^4 \ln z;\quad z=\frac{m_c}{m_b}
\end{align}
is the phase space factor in semileptonic $b$-decays. The effective Wilson coefficient 
$C_{7\gamma}^{(0)\text{eff}}(\mu)$ is given by \cite{Buras:1997fb}
\begin{align}\label{eq:c7g}
    C_{7\gamma}^{(0)\text{eff}}(\mu)=
    \eta^{16/23} C_{7\gamma}^{(0)}(M_W) 
    + \frac{8}{3} \left(\eta^{14/23}
    -\eta^{16/23}\right)\, C_{8G}^{(0)}(M_W) 
    + C_2^{(0)}(M_W)\sum_{i=1}^{8}h_i \eta^{a_i}\,,
\end{align} 
where
\beq
\eta=\frac{\alpha_s(\mw)}{\alpha_s(\mu)}\,.
\eeq
The values of $C_2^{(0)}(\mw)\sim 1$, $a_i$ and $h_i$ can be found in Ref.\ \cite{Buras:1997fb}. 
$C_{7\gamma}^{(0)}(\mw)$ and $ C_{8G}^{(0)}(\mw)$ are expressed in terms of the Inami-Lim functions 
$D'_0(x_t)$ and $E'_0(x_t)$ respectively with $x_t=\mt^2/\mw^2$:
\begin{align}
    \label{eq:c70}
    C_{7\gamma}^{(0)}(M_W)=-\frac{1}{2}D'_0(x_t)\equiv \frac{3 x_t^3 - 2x_t^2}{4(x_t -1)^4}\log{x_t} + \frac{-8 x_t^3 - 5 x_t^2 + 7 x_t}{24(x_t -1)^3}\\
    \label{eq:c80}
    C_{8G}^{(0)}(M_W)=-\frac{1}{2}E'_0(x_t) \equiv \frac{-3x_t^2}{4(x_t -1)^4}\log{x_t} + \frac{-x_t^3 + 5 x_t^2 + 2x_t}{8(x_t -1)^3} 
\end{align}
The other parameters entering in Eq.\ (\ref{eq:c7g}) are given in Table \ref{tab:sminputdb1}. 
A similar analysis follows for $b\to d\gamma$, with the suitable replacement of $s\to d$. 

\begin{table}[!htb]
\centering
 \begin{tabular}{||c|c||c|c||} 
 \hline
 Parameter & Value & Parameter & Value \\ [0.5ex] 
 \hline\hline
  $\a (M_Z)$ & $1/129$ & $m_b$ & $4.18~{\rm GeV}$\\ \hline
  $\a_s(M_Z)$ & $0.1179$ & $M_Z$ & $91.1876~{\rm GeV}$  \\ \hline
  $\sw^2$ & $0.231$ & $m_\mu$ & $105.6584~{\rm MeV}$ \\ \hline
 $\tau_{B_{sH}}$ & $1.620~{\rm ps}$  & $F_{B_s}$ & $230.3\pm1.3 ~{\rm MeV}$ \\
 \hline
\end{tabular}
\caption{Values of different parameters required for evaluating $\Delta B=1$ processes taken from Ref.\ \cite{pdg},
and $F_{B_s}$ from Ref.\ \cite{Aoki:2019cca}.}
\label{tab:sminputdb1}
\end{table}

\subsubsection{\texorpdfstring{$\bsmumu$}{Bs->mu+ mu-}}

The branching ratio for $\bsmumu$ is \cite{pdg}
\beq
{\rm Br} (\bsmumu) = \left( 3.0\pm 0.4\right)\times 10^{-9}\,,
\eeq
whereas the SM prediction is \cite{Bobeth:2013uxa,Hermann:2013kca} 
\beq
{\rm Br}(\bsmumu)_{\rm SM} = (3.65\pm 0.23)\times 10^{-9}\,.
\eeq
The expression for the branching ratio is given by
\beq
\label{eq:brbmumu}
\text{Br}(\bsmumu) = 
\tau_{B_{sH}}\, \frac{G_F^2}{\pi}\, \left(\frac{\alpha}{2\pi \sw^2} \right)^2\, F_{B_s}^2 \mmu^2 \mbs
\, \sqrt{1-4\frac{\mmu^2}{\mbs^2}}\, \left\vert V_{tb}^* V_{ts}\right\vert^2 \, \left\vert C_A(\mu_b)\right\vert^2\,,
\eeq
where $\tau_{B_{sH}}$ is the lifetime of the heavier $B_s$ eigenstate\footnote{For $B^0$, one can just 
use the average of the lifetimes of the two neutral mass eigenstates. The width difference is non-negligible 
for $B_s$.}. The Wilson coefficient $C_A$, evaluated at the regularisation scale $\mu_b$, gets contribution 
from both $W$-mediated box and $Z$-mediated penguin amplitudes:
\beq
C_A = C_{A,W} + C_{A,Z}\,,
\eeq
and each of the $C_A$s can be written as 
\beq
C_{A,W(Z)} = \sum_{n=0}^\infty C_{A,W(Z)}^{(n)} \, \left( \frac{\alpha_s}{4\pi}\right)^n\,,
\eeq
indicating the leading order (LO) terms and the higher-order QCD corrections; results up to $n=2$ can be found in 
Ref.\ \cite{Hermann:2013kca}. The LO contribution is given by the Inami-Lim function
\beq
C_A^{(0)} = \frac{x_t}{16}\, \left[ \frac{x_t -4}{x_t -1} + \frac{3x_t}{(x_t-1)^2}\, \log{x_t} \right]\,.
\eeq
The higher-order contributions can be parametrized by 
\beq
C_A(\mu_b) = \eta_Y C_A^{(0)}\,,
\eeq
with $\eta_Y = 0.905$. 

\subsection{Other loop-induced processes}
\subsubsection{\texorpdfstring{$t\to c\gamma$}{t->c gamma}}

The partial decay width for the radiative decay $t\to c\gamma$ is given by
\beq 
\label{eq:tcg}
\Gamma(t \to c\g)=\frac1\pi\, \left[ \frac{m_t^2-m_c^2}{2m_t}\right]^3 \, \left(\abs{A_\g}^2 + \abs{B_\g}^2\right)\,,
\eeq
where $A_\g$ and $B_\g$ are respectively the vector and axial form factors, whose expressions
may be found in Ref.\  \cite{AguilarSaavedra:2002ns}. 
The three-point Passarino-Veltman $C$-functions involved in the form factors are numerically evaluated 
with LoopTools \cite{Hahn:1998yk}. 
The SM prediction for ${\rm Br}(t\to c\gamma)$ is approximately $4 \times 10^{-14}$.

\subsubsection{\texorpdfstring{$t\to cZ$}{t->cZ}}

In the SM, the partial decay width for $t\to cZ$ is given by \cite{Eilam:1990zc}:
\beq
\Gamma(t \r cZ)= \Gamma(t \to c\g)+
\frac{\lambda^{1/2}(1,\,\mz^2/m_t^2,\,m_c^2/m_t^2)}{16\pi \, m_t M_Z^2}\, {\cal F}\,,
\eeq
where $\lambda(a,b,c)=a^2+b^2+c^2-2(ab+bc+ac)$, 
and ${\cal F}$ contains the CKM factors and the momenta of the particles involved. 
The branching fraction for the decay in SM is of the order of $\sim 10^{-14}$.

\subsubsection{\texorpdfstring{$\zbb$}{Z->bb}}

The ratio $R_b$, defined as
\beq \label{eq:rb}
    R_b = \frac{\Gamma(Z\to b\bar{b})} {\Gamma (Z\to {\rm hadrons})}=(1+2/R_s+1/R_c+1/R_u)^{-1}
\eeq
with $R_q \equiv \Gamma(Z\to b\bar{b})/\Gamma(Z\to q\bar{q})$, involves two non-universal corrections, namely,
$\d_{\rm tQCD}^b$, an $\mathcal{O}(\a_s^2)$ correction originating from doublets with large mass splitting, 
and $\d_b$, that represents the additional contribution to the $Zb\bar{b}$ vertex due to non-zero top quark 
mass \cite{Bernabeu:1990ws}: 
\beq \label{eq:db}
 \d_b = 2\, \frac{v_b+a_b}{v_b^2+a_b^2}\, \text{Re}\,{\d_{b\text{-vertex}}},\qquad 
 \d_{b\text{-vertex}}=2\, \left(\frac{\a}{2\pi}\right)\, \abs{V_{tb}}^2 F(x_t)
\eeq
with $x_t=(m_t/M_W)^2$ and
\begin{eqnarray}
    F(x_t) &=& \frac{1}{8 s_W^2}\, \left[ x_t + 2.880 \log x_t - 6.716 + \frac{1}{x_t}\, (8.368 \log x_t -3.408) \right. 
    \nonumber\\
    &&\left. + \frac{1}{x_t^2}\, (9.126 \log x_t + 2.260) + \frac{1}{x_t^3}\, (4.043 \log x_t +7.410) +\ldots \right]\,.
\label{fvert}    
\end{eqnarray}
Using the expressions given in Ref. \cite{Bernabeu:1990ws} for $R_c,R_s,R_u$ and $\d_{tQCD}^b$, one gets
$R_b \approx 0.217$.

\section{Modifications with vector-like quarks} \label{sec:expvlq}

With singlet or doublet VQs introduced, the processes discussed above get modified. Table \ref{tab:vqeff}
provides a quick display, which we elaborate in this Section. The model VQ-S-D1 cannot be constrained from 
any of these observables, except the row unitarity shown in Eq.\ (\ref{eq:rowunit}).
This, in conjunction with the fact that $\Delta M_d$ and $\Delta M_s$ are not affected by the VQ so that 
the SM estimates for $|V_{td}|$ and $|V_{ts}|$ are still valid, leads to a tiny value of $|V_{tb'}|$.  

\begin{table}[!htb]
\centering
 \begin{tabular}{||c||c|c|c|c|c|c||} 
 \hline
 &&&&&& \\
 Model & $\bsbsbar$ & $\bsg$ & $\bsmumu$ & $t\to c\gamma$ & $t \to cZ$ & $\zbb$  \\ 
 &&&&&& \\
 \hline\hline
 VQ-S-D1  &   ---   &   ---    &   ---   & ---    &   ---  & --- \\
 \hline
 VQ-S-D2  & $\surd$ & $\surd$ & $\surd$ & $\surd$ & $\surd$ &   ---   \\
 \hline
VQ-S-U1  & --- &  ---   &  ---   &  --- & ---  &  $\surd$  \\
\hline
 VQ-S-U2  & $\surd$ & $\surd$ & $\surd$ & $\surd$ & $\surd$ &  $\surd$   \\
\hline
 VQ-D-U2  & $\surd$ & $\surd$ & $\surd$ & --- &  ---  &  $\surd$ \\
 \hline
 VQ-D-D2 & --- & --- & --- & $\surd$ & $\surd$ & --- \\
 \hline
\end{tabular}
\caption{Effect of vector-like quarks on some flavour observables. Only for VQ-S-U2,
$t\to cZ$ occurs at the tree-level, with our ansatz for the texture of the quark mixing matrices.}
\label{tab:vqeff}
\end{table}

\subsection{VQ-S-D2} \label{sec:mod2}

\subsubsection{\texorpdfstring{$\bsbsbar$ mixing}{Bs-Bs mixing}}

Due to the presence of a new vector-like down type singlet mixing with two generations, Eq.\ (\ref{eq:delmq})
now reads  \cite{Barenboim:1997pf,AguilarSaavedra:2002kr} 
\beq
    \Delta M_s = \frac{G_F^2}{6\pi^2}\, \eta_B \mbs \left(B_{B_s}F_{B_s}^2\right)\, \mw^2 
    \left[\abs{V_{ts} V_{tb}^\ast}^2 S_0(x_t) -8 \left\vert X_{sb} V_{ts}V_{tb}^*\right\vert\, Y_0(x_t)+
    \frac{4\pi \sw^2}{\a}\, \frac{\eta_Z^{B_s}}{\eta_B}\left\vert X_{sb}\right\vert^2\right]\,,
\label{delms-vq1}
\eeq
with
\beq \label{eq:xij}
X_{ij} = \sum_{r=u,c,t}  V_{ri}^\ast V_{rj}
\eeq
and
\begin{align}\label{eq:y0}
&Y_0(x_t)=\frac{x_t}{8}\bigg[\frac{x_t-4}{x_t-1}+\frac{3x_t}{(x_t-1)^2}\log{x_t}\bigg]\,.
\end{align}
In Eq.\ (\ref{delms-vq1}), the first term inside the parenthesis gives the SM box contribution, the second term 
comes from the amplitude with a tree-level FCNC coupling on one side and the SM loop on the other, 
while the last term is a pure $Z$-mediated tree-level FCNC contribution. 
We neglect the contributions due to charm and up quarks, and take $\eta_Z^{B_s}=0.57$ 
\cite{AguilarSaavedra:2002kr}.

\subsubsection{\texorpdfstring{$\bsg$}{b->s gamma}}

With a charge $-\frac13$ VQ $b'$ introducing FCNC couplings in the down-quark sector, the radiative decay 
$\bsg$ involves more penguin amplitudes, where the canonical $W$-propagator is replaced by either the 
$Z$ boson, or the Higgs boson $h$ \cite{Chang:2000gz}. The additional contributions to the Wilson 
coefficients $C_{7\gamma}$ and $C_{8G}$ in Eqs.\ (\ref{eq:c70}) and (\ref{eq:c80}) are given by
\bea
 \delta C_{7\gamma}(\mw) &=& 
 \frac{X_{sb}}{\lamsb}\, \left( \frac{23}{36}+\xi_s^Z +\xi_b^Z\right) + \frac{X_{sb'} X_{b'b}}{\lamsb}\, 
 \left[\xi_{b'}^Z(y_{b'}) +\xi_{b'}^h(w_{b'})\right]\,,\nonumber\\
 \delta C_{8G}(\mw) &=& \frac{X_{sb}}{\lamsb}\, \left(\frac{1}{3}-3\xi_s^Z -3\xi_b^Z\right) - 
 3 \frac{X_{sb'} X_{b'b}}{\lamsb}\, \left[\xi_{b'}^Z(y_{b'}) +\xi_{b'}^h(w_{b'})\right]\,,
\eea
where $X_{sb'}X_{b'b} \approx -X_{sb}$ (assuming $\abs{{\cal D}_{44}}^2 \approx 1$), and
\bea
\xi_s^Z &=& \frac{1}{54}\, \left(-3+2 \sw^2\right)\,, \quad \xi_b^Z=\frac{1}{54}\, \left(-3-4 \sw^2\right)\,,\nonumber\\
\xi_{b'}^Z (y_{b'}) &=& -\frac{8-30y_{b'} +9y_{b'}^2 -5y_{b'}^3} {144(1-y_{b'})^3} + \frac{y_{b'}^2}{8(1-y_{b'})^4}\log{y_{b'}}\,,\nonumber\\
\xi_{b'}^h (w_{b'}) &=& -\frac{16w_{b'} -29w_{b'}^2 +7w_{b'}^3} {144(1-w_{b'})^3} + 
\frac{-2w_{b'}+3w_{b'}^2}{24(1-w_{b'})^4}\log{w_{b'}}\,,
\eea
with $y_i=(m_i/M_Z)^2$ and $(w_i=m_i/M_h)^2$ \cite{Chang:2000gz}. 
Three other Wilson coefficients, zero in the SM, receive non-zero contribution in this scenario:
\beq
    C_3(\mw) = -\frac{1}{6}\frac{X_{sb}}{\lamsb}\,, \quad C_7(\mw) = -\frac{2}{3}\sw^2 \frac{X_{sb}}{\lamsb}\,, 
    \quad C_9(M_W) = \frac{2}{3}(1-\sw^2)\frac{X_{sb}}{\lamsb}\,.
\eeq
At the scale $\mu=5$ GeV, one gets
\bea
    C_{7\gamma}^{(0)\text{eff}}(\mu) &=& -0.158 C_2(\mw) + 0.695 C_{7\gamma}(\mw) + 0.085 C_{8G}(\mw)\n\\ 
    && +0.143 C_3(\mw) + 0.101 C_7(\mw) -0.036 C_9 (\mw)\,.
\eea
In the SM, $C_2(\mw)=1$ because of the unitarity of the CKM matrix; here, it is better to use 
\beq
C_2(\mw)=-\frac{V_{cs}^\ast V_{cb}}{V_{ts}^\ast V_{tb}}\,.
\eeq
 
\subsubsection{\texorpdfstring{$\bsmumu$}{Bs->mu+ mu-}}

The contribution due to tree level $Z$ exchange to Eq. (\ref{eq:brbmumu}) is given by \cite{Barger:1995dd}
\beq
{\rm Br}(\bsmumu)_{\rm VQ} = \tau_{B_{sH}}\frac{G_F^2 F_{B_s}^2 \mbs m_{\mu}^2} {8\pi} \, 
\sqrt{1-\frac{4m_\mu^2}{\mbs^2}}\, \left[\left(\frac12 - \sw^2\right)^2 +\sw^4 \right]\, |X_{sb}|^2\,.
\eeq

\subsubsection{\texorpdfstring{$t \to c\g$}{t->c gamma}}

The loop mediated contributions of $b'$ to the form factors in Eq.\ (\ref{eq:tcg}) can be easily incorporated as
\begin{align}
    \d\, A_\g = A_{\g,i}^{b'}, \quad  \d\, B_\g = B_{\g,i}^{b'}
\end{align}
where $i$ indicates the contribution coming from various diagrams \cite{AguilarSaavedra:2002ns}.

\subsection{VQ-S-U1} \label{sec:mod3}

With an additional up-type singlet quark $t'$ and only $V_{t'b}\not=0$, a similar constraint as for VQ-S-D1 
comes from the column unitarity $\sum |V_{ib}|^2 = 1$. However, there is one decay channel where $t'$ can 
affect. This is the modification of the $Zb\bar{b}$ vertex; the loop corrections due to $t$ and $t'$ 
modify the decay width $\Gamma(Z \to b \bar{b})$ and thus $R_b$. As all the effects due to large top quark mass are contained in the vertex correction factor $\d_b$, the mixing of top quark with vector quarks modifies the function $F$ in Eq. (\ref{eq:db}), which is taken into account by making the substitution $F \r F+F_2$ \cite{AguilarSaavedra:2002kr,Bamert:1996px}, where
\beq
F_2(x_t)=\frac{1}{8 s_W^2}\frac{Z_{tt}-1}{2}x_t \bigg(2-\frac{4}{x_t-1}\log x_t \bigg)\n
\eeq
and
\beq
Z_{ij} = \sum_{r=d,s,b}  V_{ir}^\ast V_{jr}\,. \label{eq:zij}
\eeq
In addition to diagrams involving top quarks in the loop, there are triangle diagrams involving $t'$s in the loop, or $t$ 
and $t'$. Therefore, $\d_b$ is modified as    
\begin{align}\label{eq:dbvq}
    \d_{b\text{-vertex}} \approx \frac{\a}{\pi}\, \left(\abs{V_{tb}}^2 \left[ F(x_t)+F_2(x_t)\right]
    +\abs{V_{t'b}}^2 \left[ F(x_{t'})+F_2(x_{t'})\right]+V_{tb}^\ast V_{t'b}(x_t,x_{t'})F_3(x_t,x_{t'})\right)\,,
\end{align}
where the $t \hyp t'$ contribution is given by the last term \cite{AguilarSaavedra:2002kr}, with
\begin{align}    
    \begin{split}
     F_3(x_t,x_{t'})&=\frac{1}{2s_W^2}\frac{\text{Re}Z_{tt'}}{2}\bigg[-\frac{1}{x_t-x_{t'}}\bigg(\frac{x_{t'}^2}{x_{t'}-1}\log x_{t'}-\frac{x_{t}^2}{x_{t}-1}\log x_{t}\bigg)\\
     &+\frac{x_t x_{t'}}{x_{t'}-x_t}\bigg(\frac{x_{t'}}{x_{t'}-1}\log x_{t'}-\frac{x_{t}}{x_{t}-1}\log x_{t}\bigg)
     \bigg]\,.   
    \end{split}
\end{align}
Since, $t'$ mixes only with third generation in this model, we have $Z_{tt'}=V_{tb}^\ast V_{t'b}\,.$

\subsection{VQ-S-U2} \label{sec:mod4}

\subsubsection{\texorpdfstring{$\bsbsbar$ mixing}{Bs-Bs mixing}}

With the VQ $t'$ in the loop, Eq. (\ref{eq:delmq}) is modified as
\bea
    \Delta M_s &=&
    \frac{G_F^2}{6\pi^2} \mbs (B_{B_s}F_{B_s}^2)\mw^2 \left[\eta_{tt}\, \abs{V_{ts} V_{tb}^\ast}^2 S_0(x_t)
    + \eta_{t't'}\, \abs{V_{t's} V_{t'b}^\ast}^2 S_0(x_{t'}) \right.\nonumber\\
    && \left. +2 \eta_{tt'} V_{ts} V_{tb}^\ast \,  V_{t's} V_{t'b}^\ast \tilde{S}_0 (x_t,x_{t'})
    \right]\,,
\eea
where
\beq
\tilde{S}_0(x_t,x_{t'}) = x_t\, \left[ \log \frac{x_{t'}}{x_t}-\frac{3x_{t'}}{4(1-x_{t'})}-\frac{3x_{t'}^2\, \log{x_{t'}}}
{4(1-x_{t'}^2)} \right]\,.
\eeq
In our analysis, we will take all QCD correction factors to be equal:
\beq
\eta_{tt} \approx \eta_{t't'} \approx \eta_{tt'} = \eta_B\,,
\eeq
and neglect the contribution coming from the up and the charm quarks.

\subsubsection{\texorpdfstring{$\bsg$}{b->s gamma}}

Similarly, for the radiative decay with the $t'$ quark loop, the additional contribution to the Wilson coefficients in 
Eqs.\ (\ref{eq:c70}) and (\ref{eq:c80}) can be written as
\begin{align}
    \d C_{7\g}(M_W)=-\frac{1}{2}\, \frac{V_{t's}^{*}V_{t'b}}{V_{ts}^{*}V_{tb}} D'_0(x_{t'})\,,\nonumber\\
    \d C_{8G}(M_W)=-\frac{1}{2}\, \frac{V_{t's}^{*}V_{t'b}}{V_{ts}^{*}V_{tb}} E'_0(x_{t'})\,,
\end{align}
where
\begin{align}
D'_0(x_{t'})= \frac{-8 x_{t'}^3 - 5 x_{t'}^2 + 7 x_{t'}}{12(1-x_{t'})^3}+\frac{-3 x_{t'}^3 + 2x_{t'}^2}{2(1-x_{t'})^4}\log{x_{t'}}\,,\nonumber \\
E'_0(x_{t'})= \frac{-x_{t'}^3 + 5 x_{t'}^2 + 2x_{t'}}{4(1-x_{t'})^3}+\frac{3x_{t'}^2}{2(1-x_{t'})^4}\log{x_{t'}}\,.
\end{align}

\subsubsection{\texorpdfstring{$\bsmumu$}{Bs->mu+ mu-}}

In presence of $t'$, the branching ratio for $\bsmumu$ is given by
\begin{align}\label{eq:45}  
    {\rm Br}(\bsmumu)=\tau_{B_{sH}}\, \frac{G_F^2}{\pi}\left( \frac{\alpha}{4\pi \sw^2} \right)\, 
    F_{B_s}^2 m_\mu^2 \mbs \sqrt{1-4\frac{m_\mu^2}{\mbs^2}}\, Y_m^2 \,,
\end{align}
where
\begin{align}
    Y_m=\eta_t V_{tb}^{*}V_{ts} Y_0(x_t) + \eta_{t'} V_{t'b}^{*}V_{t's} Y_0(x_t')\,.
\end{align}
Again, we use the approximation $\eta_t \approx \eta_{t'}=\eta_Y$ for our estimates.

\subsubsection{\texorpdfstring{$t \to c\g$}{t->c gamma}}

The introduction of $t'$ induces new FCNC vertices in the up-quark sector. Thus, there will be new loop
amplitudes with FCNC $Z$-vertices. 
The flavour diagonal couplings also get modified according to Eq.\ (\ref{eq:fcnc3}). The contribution of the new diagrams to the form factors in Eq.\ (\ref{eq:tcg}) is given in Ref.\ \cite{AguilarSaavedra:2002ns}.

\subsubsection{\texorpdfstring{$t \to cZ$}{t->cZ}}

Only in this class of models the FCNC decay $t\to cZ$ occurs at the tree-level, 
and its decay width, with next-to-leading order QCD corrections, can be written as
\beq
\Gamma(t \r cZ)=\frac{g^2}{128\pi\, c_W^2}\,\abs{Z_{ct}}^2\,\frac{m_t^3}{M_Z^2}\left(1-\frac{M_Z^2}{m_t^2}\right)^2 \, \left(1+2\frac{M_Z^2}{m_t^2} \right)\, \left[1-\frac{2\a_s}{3\pi}\left(\frac{2\pi^2}{3}-\frac{5}{2}\right)  \right]\,,
\eeq
where the charm quark mass has been neglected.

\subsubsection{\texorpdfstring{$\zbb$}{Z->bb}}

The contribution of the new singlet to the vertex correction factor for $\zbb$ is described in Section \ref{sec:mod3}. However, as $t'$ mixes with two generations in this model, from Eq.\ (\ref{eq:zij}) we have
\beq
Z_{tt'}=V_{ts}^\ast V_{t's}+V_{tb}^\ast V_{t'b}\,.
\eeq

\subsection{VQ-D-U2} \label{sec:mod5}

In this model, the processes $\bsbsbar$ mixing, $\bsg$ and $\bsmumu$ receive an additional contribution 
from $t'$ in the loop similar to the model VQ-S-U2, and the corresponding expressions in Section \ref{sec:mod4}
can be used. However, since there is no tree level FCNC in the left-chiral as well as the right-chiral sector (our
choice of ${\cal U}_R=\mathbf{1}$ ensures this), the process $t \to c\g$ does not receive any contribution 
from the new quarks. For $Z \to b\bar{b}$, the vertex correction factor of Eq.\ (\ref{eq:db}) can be written as
\beq
  \d_{b\text{-vertex}}=\frac{\a}{\pi}\, \left(\abs{V_{tb}}^2 F(x_t) + \abs{V_{t'b}}^2 F(x_{t'}) \right)\,.
\eeq
Apart from these processes, one must also consider the constraints coming from the unitarity of 
the $4\times 4$ CKM matrix, as in Eq.\ (\ref{eq:unitarity}).

\subsection{VQ-D-D2} \label{sec:mod6}

Here, as all elements of the fourth row except $V_{t'b'}$ are taken to be zero, the only affected process 
is $t \to c \g$, originating from the contribution of $b'$ at one loop. There are no effects due to $t'$, as FCNC
couplings are absent. The contribution to the form factors are, therefore, same as that shown in Section
\ref{sec:mod2}.

\section{Results} \label{sec:results}

Let us now discuss the bounds on the elements of the quark mixing matrix, in the presence of one or more
VQs. To be on the conservative side, we take the theoretical uncertainties as well as experimental error margins
at $2\sigma$, {\em i.e.}, 95\% confidence level (assuming the uncertainties to be Gaussian in nature).
The benchmark mass for all the VQs is taken to be 2 TeV, just outside the LHC detection limit.  

The one-loop $\Delta B = 1$ and $2$ processes, namely, $\bsg$, $\bsmumu$, and $\bsbsbar$ mixing, put 
bounds on $V_{ts}$, $X_{sb}$ (Eq.\ (\ref{eq:xij})), and  ${\lambda_{sb}^{t'}}\equiv \lamsbtp$. 
In view of the importance of QCD corrections, and uncertainties associated with fixing $\mu_{\text{eff}}$ 
for the radiative decay $\bsg$, we just compare the branching ratios in the VQ model vis-\`a-vis the SM. This 
is an approximation on the universality of the higher-order effects for the SM quarks and the VQs, but at 
least holds for the QCD corrections. Thus, we have  
\begin{align}
    \frac{\text{Br}(B\r X_s \g)}{\text{Br}(B\r X_s \g)_{\text{SM}}}=
    \left(\frac{\abs{V_{ts}^\ast V_{tb}}}{\abs{V_{ts}^\ast V_{tb}}_{\text{\tiny{SM}}}}\right)^2 \,\, \left\vert \frac{C_{7\g}^{(0)\text{eff}}(\mu)}{C_{7\g \text{\tiny{SM}}}^{(0)\text{eff}}(\mu)} \right\vert^2\,,
\end{align}
where $C_{7\g \text{\tiny{SM}}}^{(0)\text{eff}}(\mu)$ is given by Eq.\ (\ref{eq:c7g}).
Similarly, to analyse the bound on ${V_{t'b}}$ from the process $Z \to b\bar{b}$, we take the ratio
\begin{align}\label{eq:rbanalysis}
    \frac{R_b}{R_b^{\text{\tiny{SM}}}} \approx \frac{1+\d_b}{1+\d_b^{\text{\tiny{SM}}}}
\end{align}
as the dominant contribution from the VQs appears in the vertex correction factor of Eq.\ (\ref{eq:db}). 

The new mixing matrix elements may contain nontrivial phases. For our analysis, we parametrize 
\beq
X_{sb}=|X_{sb}|\, \exp{(i\theta)}\,,\ \ \lsbtp=|\lsbtp|\, \exp{(i\delta)}\,.
\eeq
For the complex phase, we use the measurement of $\beta_s$ in the $\bsbsbar$ system:
\begin{align}
	M_{12}^s =\abs{M_{12}^s}\, \exp(-2i\beta_s)
\end{align} 
with $\beta_s=(1.1 \pm 1.6) \times 10^{-2}$.

In a similar vein, the bounds on $V_{cb'}$, $V_{tb'}$, and ${Z_{ct}}$ can be obtained from $t \to c\g$. 
However, this is not yet observed and only an upper limit, orders of magnitude above the SM prediction,
exists \cite{pdg}. Thus, no stringent bounds on these elements can be obtained from $t\to c\g$;
rather, constraints from unitarity seem more promising.

The SM estimates and experimental numbers for the observables have been listed in Table \ref{tab:smexp}. 
Note that the SM numbers for $\bsg$ or $\bsmumu$ involve $V_{ts}$ which, in turn, is obtained from the 
$\bsbsbar$ system. We discuss the bounds on the magnitude and phase of the new mixing matrix elements 
for different VQ models, and summarise our results in Table \ref{tab:constckm}. For all the VQ types, 
we generate several models, specified by the new mixing matrix elements, and see if they pass all the 
experimental constraints. This produces a scatter plot for the allowed parameter space. 

\begin{table}[!htb]
\centering
 \begin{tabular}{||c||c|c|c||} 
 \hline
  Observables & Measurement & SM Prediction & Reference \\ [0.5ex] 
 \hline\hline
  $R_b$ & $0.21629 \pm 0.00132$ & $0.21578 \pm 0.00022$ & \cite{Gori:2015nqa,pdg} \\
  \hline
  $\text{Br}(b \r s\g)$ & $(3.32 \pm 0.30) \times 10^{-4}$ & $(3.36 \pm 0.46) \times 10^{-4}$ 
  & \cite{Misiak:2015xwa,Misiak:2017woa}\\
  \hline 
  $\text{Br}(\bsmumu)$ & $\left(3.0\pm 0.8\right) \times 10^{-9}$ & $(3.65 \pm 0.46) \times 10^{-9}$ &
   \cite{Bobeth:2013uxa,pdg}\\ [0.3ex]
  \hline 
  $\text{Br}(t\r c\g)$ & $<1.8 \times 10^{-4}$ & $\sim 4 \times 10^{-14}$ & \cite{pdg}\\
  \hline
 $\text{Br}(t\r cZ)$ & $<5 \times 10^{-4}$ & $\sim 1 \times 10^{-14}$ & \cite{pdg,AguilarSaavedra:2004wm}\\
  \hline
\end{tabular}
\caption{SM estimates and experimental numbers for several observables with $2\sigma$ error margin.}
\label{tab:smexp}
\end{table}

\subsection{VQ-S-D1}

The only constraint comes from Eq.\ (\ref{eq:rowunit}), assuming ${\cal U}={\bf 1}$:
\beq
\left\vert V_{tb'}\right\vert \leq 0.30\,,
\eeq
which results in a fast decay $b'\to tW$.

\subsection{VQ-S-D2}

\begin{figure}[!htb]
\centering
\includegraphics[scale=0.4]{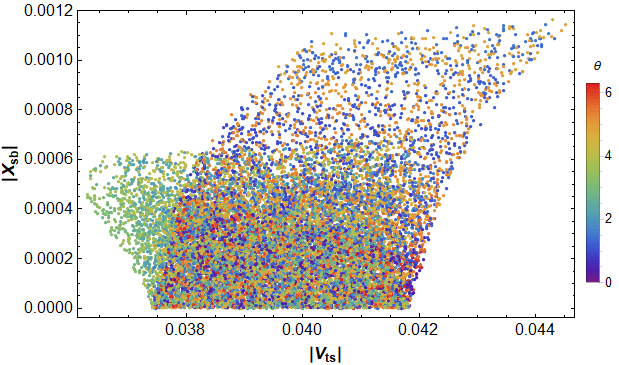} \ \ 
\includegraphics[scale=0.3]{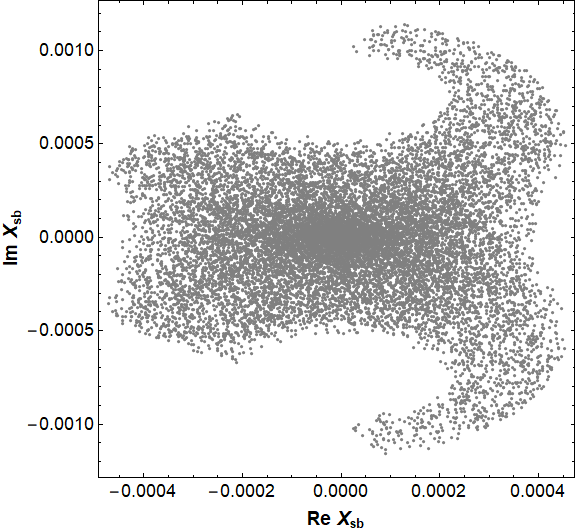}
\caption{Constraints from $\Delta B=1$ and $\Delta B=2$ processes on 
$\abs{V_{ts}}$ and $X_{sb}$ for the model VQ-S-D2.}
\label{fig:vq-s-d2}
\end{figure}

The $\Delta B=1$ and $\Delta B = 2$ processes put bound on $X_{sb}$ as well as $\abs{V_{ts}}$, which 
are shown in Figure \ref{fig:vq-s-d2}. Combined constraints from all the three processes, namely, 
$\bsg$, $\bsmumu$, and $\bsbsbar$ mixing, yield $|X_{sb}|<1.2 \times 10^{-3}$ and 
$0.036 < \abs{V_{ts}} < 0.045$. One may note the role the phase $\theta$, associated with $X_{sb}$,
plays in determining the allowed parameter space for both $\abs{X_{sb}}$  and $\abs{V_{ts}}$. 
On the other hand, constraints from the row unitarity of the $3 \times 4$ CKM matrix, 
Eq.\ (\ref{eq:rowunit}), give $|V_{cb'}|\leq 0.15$ 
and $|V_{tb'}|\leq 0.30$. Using the above bounds, the maximum branching ratio for $t\to c\g$ comes out 
to be $\sim \mathcal{O}(10^{-8})$, still well below the current reach. 
     
\subsection{VQ-S-U1}

The only process where this model contributes is $\zbb$, and gives a bound on $V_{t'b}$. The minimum 
value of $Z_{tt} = \abs{V_{td}}^2+\abs{V_{ts}}^2+\abs{V_{tb}}^2$, which is a measure of the deviation from 
unitarity of the third row of the CKM matrix, comes out to be $Z_{tt}\approx 0.91$.
From Eqs.\ (\ref{eq:db}), (\ref{eq:dbvq}), and (\ref{eq:rbanalysis}), we obtain $\abs{V_{t'b}} < 0.047$.
The column unitarity produces a worse bound: $\abs{V_{t'b}} < 0.30$. 
	
\subsection{VQ-S-U2 and VQ-D-U2}

\begin{figure}[!htb]
\centering
\includegraphics[scale=0.43]{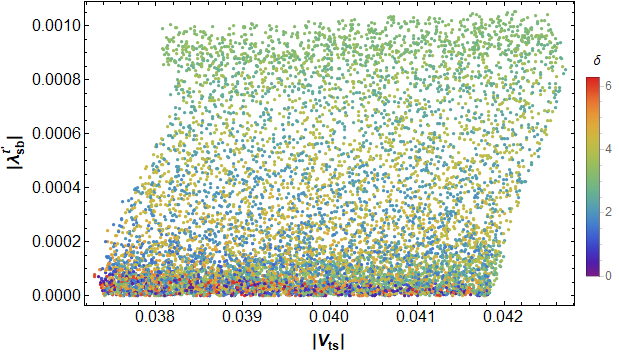} \ \ 
\includegraphics[scale=0.35]{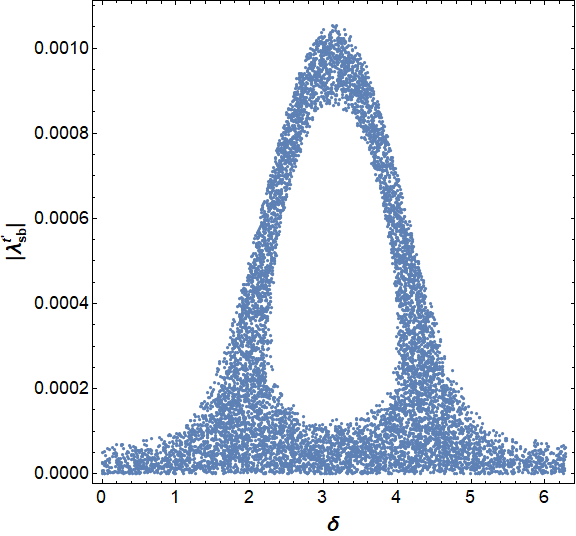}
\caption{Constraints from $\Delta B=1$ and $\Delta B=2$ processes on $\abs{V_{ts}}$ and $\lambda_{sb}^{t'}$ 
for the models VQ-S-U2 and VQ-D-U2.}
\label{fig:vq-s-u2}
\end{figure}

Most of the affected processes and the bounds obtained are identical for these two models, so we lump 
them together. In Fig.\ \ref{fig:vq-s-u2}, we show the bounds on ${\lambda_{sb}^{t'}}$ and $\abs{V_{ts}}$
 arising from $\bsbsbar$ mixing, $\bsg$ and $\bsmumu$. The hollow bell-shaped plot, the right-hand panel of 
 Fig.\ \ref{fig:vq-s-u2}, is interesting. Large values of $\lambda_{sb}^{t'}$ near $\delta = \pi$ indicate a 
 destructive interference with the SM, with the VQ amplitude about twice in magnitude compared to the SM one. 
 
Combined constraints from all the three processes give $\abs{\lambda_{sb}^{t'}}<1.1 \times 10^{-3}$ and 
$0.037 < \abs{V_{ts}} < 0.043$. 
Constraints from the column unitarity of the $4 \times 3$ CKM matrix give $\abs{V_{t's}} \leq 0.13$ and 
$\abs{V_{t'b}} \leq 0.30$. However, one gets a better bound from $\zbb$, keeping $\abs{V_{t's}} \sim 0.13$: 
$\abs{V_{t'b}} < 0.042$. This worsens slightly to $\abs{V_{t'b}}  <  0.045$ for the VQ-D-U2 model.  

From $t\to cZ$ (only for VQ-S-U2), one obtains $\abs{Z_{ct}}<0.031$. With $\abs{Z_{ct}} \sim 
\mathcal{O}(10^{-2})$,  ${\rm Br}(t \to c\g)$ comes out to be  $\sim \mathcal{O}(10^{-8})$. 
For the texture of the CKM matrix that we have chosen, there is no such FCNC for VQ-D-U2.

\subsection {VQ-D-D2}

The treatment for the only affected process, namely, $t\to c\g$, is identical to that of VQ-S-D2. 

A brief summary of our results is displayed in Table \ref{tab:constckm}. We also show, in Table
\ref{tab:1.5-10tev}, the effect of the VQ mass on these constraints, starting from the present lower bound
of $1.5$ TeV (approximately) and going up to 10 TeV. Because of the behaviour of the Inami-Lim functions, 
we find, as expected, that the VQs decouple from the SM (the upper bound on the mixing becomes smaller)
with increasing mass.

\begin{table}[!htb]
\centering
  \begin{tabular}{||c||c|c|c|c|c|c|c||}
\hline
  Model & $|V_{ts}|$ & $|X_{sb}|$ & $|Z_{ct}|$ & $|\lsbtp|$ & $|V_{t'b}|$ & $|V_{cb'}|$ & $|V_{tb'}|$\\ [0.5ex] 
 \hline\hline
 	VQ-S-D1 & --- & 0 & --- & --- & --- & $\leq 0.15^\dag$ & $\leq 0.30^\dag$ \\
  \hline
  VQ-S-D2 & 0.036 -- 0.045 & $< 0.0012$ & --- & --- & --- & $\leq 0.15^\dag$ & $\leq 0.30^\dag$ \\
  \hline
  VQ-S-U1 & --- & --- & 0 & 0 & $<0.047$ & --- & ---  \\
  \hline
  VQ-S-U2 & 0.037 -- 0.043 & --- & $<0.031$ & $< 0.0011$ & $<0.042$ & --- & ---  \\
  \hline
   VQ-D-U2 & 0.037 -- 0.043 & --- & --- & $< 0.0011$ & $<0.045 $ & 0 & 0 \\
  \hline 
  VQ-D-D2 & --- & --- & --- & 0 & 0 & $\leq 0.15^\dag$ & $\leq 0.30^\dag$ \\
  \hline 
\end{tabular}
\caption{Constraints on the quark mixing matrix elements and their combinations in presence of 
vector-like quarks. A dash indicates that these observables are not affected in the model, whereas 
entries marked with a dagger indicate that they have been obtained from unitarity constraints. Note that 
we have taken $V_{t'd},V_{ub'}=0$ for all the models.}
\label{tab:constckm}
\end{table}

\begin{table}[!htb]
\centering
\begin{tabular}{||c|| *{3}{c|}c|| }
    \hline
    & \multicolumn{2}{c|}{$|\lsbtp|$}
                            & \multicolumn{2}{c||}{$|V_{t'b}|$} \\
    \hline
$m_{t',b'}$  &   1.5 TeV   &   10 TeV   &   1.5 TeV  &   10 TeV  \\
    \hline \hline
  VQ-S-U1 & \multicolumn{2}{c|}{0} & $<0.062$ & $< 9.8\times 10^{-3}$ \\
  \hline
  VQ-S-U2 &  $1.9\times 10^{-3}$ & $4.4\times 10^{-5}$ & $<0.054$ & $< 9.3 \times 10^{-3}$ \\
  \hline
  VQ-D-U2 & $1.9\times 10^{-3}$ & $4.4\times 10^{-5}$ & $<0.060$ & $<9.1 \times 10^{-3}$ \\
  \hline   
\end{tabular}
\caption{Constraints on the quark mixing matrix elements and their combinations for the VQ mass 
of 1.5 TeV and 10 TeV. Other constraints are not significantly affected.}
\label{tab:1.5-10tev}
\end{table}

\section{Summary} \label{sec:summary}

In this paper, we have considered several models with vector-like quarks, including $SU(2)$ singlet and 
doublet representations, and of charges $+\frac13$ and/or $-\frac23$, so that they can mix with their SM 
counterparts. To make our life simple, we assume that they mix only with the third generation quarks, or at the 
most, with the second and the third generation quarks. This introduces new complex elements in the 
expanded CKM matrix. At the same time, the $3\times 3$ SM block of the full CKM matrix no longer 
remains unitary. 

We use the low-energy observables, namely, $\bsbsbar$ mixing, the decays $\bsg$ and $\bsmumu$, and the 
partial decay width $R_b$, to constrain these new elements. Not all models can be constrained from these 
observables, and further extensions (like mixing with the first generation quarks) will bring in other 
observables. Our constraints have been discussed in the previous Section, which are consistent with other 
similar studies in the literature. We have also found that the width of the anomalous top decay $t\to c\g$ may 
be significantly enhanced in the presence of such VQs, by a few orders of magnitude compared to the SM, but
still remains well below the present LHC reach. The parameter space will naturally get more squeezed with 
new data from LHCb and Belle-II.

\noindent {\em Acknowledgements} --- A.K.\ acknowledges the support from the Science and Engineering Research 
Board, Govt.\ of India, through the grants CRG/2019/000362, MTR/2019/000066, and DIA/2018/000003.

\bibliographystyle{utphys}
\bibliography{refs}

\end{document}